\documentstyle[11pt]{article}
\begin{document}
\begin{titlepage}
\vspace*{-1cm}
\noindent
\phantom{bla}
\hfill{UMHEP-389}
\\
\vskip 2.5cm
\begin{center}
{\Large\bf Chiral Sum Rules and Their Phenomenology}
\end{center}
\vskip 1.5cm
\begin{center}
{\large John F. Donoghue and Eugene Golowich}
\\
\vskip .3cm
Department of Physics and Astronomy \\
University of Massachusetts \\
Amherst MA 01003 USA\\
\vskip .3cm
\end{center}
\vskip 2cm
\begin{abstract}
\noindent
We present an analysis of four sum rules, each based on chiral
symmetry and containing the difference
$\rho_{\rm V}(s) - \rho_{\rm A}(s)$ of isovector
vector and axialvector spectral functions.  Experimental data
from tau lepton decay and electron-positron scattering
identify the spectral functions over a limited kinematic domain.
We summarize the status of the existing database.  However,
a successful determination of the sum rules requires
additional content, in the form of theoretical input.
We show how chiral symmetry and the operator product expansion
can be used to constrain the spectral functions in the
low energy and the high energy limits and proceed to perform a
phenomenological test of the sum rules.
\end{abstract}
\vfill
\end{titlepage}
\vskip2truecm
\section{\bf Motivation}

Despite a concerted effort by physicists extending over many years,
an understanding of $QCD$ from first principles continues to be
elusive.  Fortunately, data continue to appear which provide a
rather direct probe of the inner workings of the strong interactions.

A case in point involves semileptonic tau lepton decay and hadron
production in $e^+e^-$ scattering.  For both of these,
multihadron states such as $2\pi ,~ 3\pi , \dots$ are excited
from the operation of quark vector and axialvector currents on the
$QCD$ vacuum.  In this paper, we shall restrict our discussion to the
isospin currents,$^{\cite{dgh}}$
\begin{equation}
V^\mu_a = {\bar q}{\tau_a \over 2}\gamma^\mu q \qquad {\rm and} \qquad
A^\mu_a = {\bar q}{\tau_a \over 2}\gamma^\mu \gamma_5 q
\ \ ,
\label{1}
\end{equation}

\noindent where $a = 1,2,3$ and $q = (u\ d)$.  Such current-induced processes
provide information about the current bilinears,
\begin{equation}
\langle 0|T\left( V^\mu_a (x) V^\nu_b (0)\right)|0\rangle \qquad {\rm and}
\qquad \langle 0|T\left( A^\mu_a (x) A^\nu_b (0)\right)|0\rangle \ \ .
\end{equation}
It has been long recognized that these quantities appear in certain
chiral sum rules.  Since these sum rules follow rather directly from
$QCD$ and chiral symmetry, a test of their correctness is, in effect,
an experimental check on the validity of $QCD$ itself.

Unfortunately, there are several formidable obstacles to
a successful implementation of this
procedure.  For one, the sum rules encompass an infinite
range of energy, whereas existing data covers a very modest
range, $s<m_\tau^2$ for tau decay and $s\le 5$~GeV$^2$
in $e^+e^-$ scattering.  Moreover, as we shall see there are
uncertainties in existing data which future experimental work
must clear up.

We feel that such difficulties can be overcome.  In this work,
we shall argue that a combination of chiral symmetry and $QCD$
sum rule methods constrain the low-energy and high-energy limits
of the dispersion integrals and that data can be used to fill in
much of the rest.  The content of the paper is organized as follows.
We begin by introducing the spectral functions and their sum rules in
Sect.~2, review the state of existing data in
Sect.~3 and then describe various theoretical constraints in Sect.~4.
Armed with experimental data and theoretical
constraints, we present details of a phenomenological analysis of
the chiral sum rules in Sect.~5.  The final part in Sect.~6
summarizes our findings.
\section{\bf Spectral Representations}

In this section, we shall work exclusively
in the chiral world of massless $u,d$ quarks.
Here, the spin-$0$ axial contribution is given by the pion pole
and the two-current time-ordered products
can be expressed in terms of spin-one spectral functions
$\rho_{\rm V,A}(s)$,$^{\cite{spec}}$
\begin{eqnarray}
\lefteqn{\langle 0|T\left( V^\mu_a (x) V^\nu_b (0)\right)
|0\rangle =} \nonumber \\
& & i\delta_{ab} \int_0^\infty ds~ \rho_{\rm V} (s)~(-sg^{\mu\nu}
- \partial^\mu \partial^\nu) \int {d^4p\over (2\pi)^4}~{e^{-ip\cdot x}\over
p^2 - s + i\epsilon}
\label{4a}
\end{eqnarray}
and
\begin{eqnarray}
\lefteqn{\langle 0|T\left( A^\mu_a (x) A^\nu_b (0)\right)|0\rangle =
- i\delta_{ab}F_\pi^2 \partial^\mu\partial^\nu
\int {d^4p\over (2\pi)^4}~{e^{-ip\cdot x}\over
p^2 + i\epsilon}}  \nonumber \\
& &  + i\delta_{ab}
\int_0^\infty ds ~\rho_{\rm A} (s)(-sg^{\mu\nu} -
\partial^\mu\partial^\nu) \int {d^4p\over (2\pi)^4}~{e^{-ip\cdot x}\over
p^2 - s + i\epsilon} \ .\label{4b}
\end{eqnarray}
For completeness, we also give the corresponding relations involving
non time-ordered products,
\begin{eqnarray}
{1\over 2\pi} \int d^4 x ~e^{iq\cdot x}
\langle 0|V^\mu_a (x) V^\nu_b (0) |0\rangle &=& i\delta_{ab}~
\rho_{\rm V} (q^2)~(q^\mu q^\nu - q^2 g^{\mu\nu} ) \nonumber \\
{1\over 2\pi} \int d^4 x ~e^{iq\cdot x}
\langle 0| A^\mu_a (x) A^\nu_b (0)|0\rangle &=& i\delta_{ab}~
[\rho_{\rm A} (q^2)~(q^\mu q^\nu - q^2 g^{\mu\nu}) \label{non} \\
&\phantom{x}& \phantom{xxx} + F_\pi^2 \delta (q^2) q^\mu q^\nu ]\ \ . \nonumber
\end{eqnarray}

It follows from chiral symmetry and the high energy behavior of $QCD$ that
the vector and axialvector spectral functions contribute to
certain sum rules,$^{\cite{anat}}$
\vspace{0.15cm}
\begin{eqnarray}
\int_{0}^\infty ds ~{\rho_{\rm V}(s) - \rho_{\rm A}(s)\over s}
&=& -4{\bar L}_{10} = (2.73 \pm 0.12 ) \times 10^{-2}
\ , \label{6} \\
\int_{0}^\infty ds ~(\rho_{\rm V}(s) - \rho_{\rm A}(s))
&=& F_\pi^2 \label{7} \\
&=& (8.54 \pm 0.06)\times 10^{-3}~{\rm GeV}^2
\ \ ,\nonumber \\
\int_{0}^\infty ds \ s~(\rho_{\rm V}(s)
- \rho_{\rm A}(s)) &=& 0 \ \ , \label{8} \\
\int_{0}^\infty ds ~s\ln\left({s\over\Lambda^2}\right)~
(\rho_{\rm V}(s) - \rho_{\rm A}(s)) &=&
-{16\pi^2 F_\pi^2\over 3e^2}  (m^2_{\pi^\pm} - m^2_{\pi^0} ) \label{9} \\
&=& -(6.19 \pm 0.03  )\times 10^{-3}~{\rm GeV}^4 ~.  \nonumber
\end{eqnarray}
\vspace{0.15cm}
In the remainder of the paper, we shall refer to the above sum rules
respectively as $W0$, $W1$, $W2$ and $W3$.  In the first one,
$W0$, the quantity ${\bar L}_{10}$ is related to
the renormalized coefficient $L_{10}^{(r)}(\mu)$ of
an ${\cal O}(E^4)$ operator appearing in the effective chiral lagrangian
of $QCD$.$^{\cite{kfsr}}$  Although the value of $L_{10}^{(r)}(\mu)$, which is
measured in the radiative pion decay $\pi \to e\nu \gamma$, refers to a
renormalization scale $\mu$, the quantity ${\bar L}_{10}$ is itself
independent of $\mu$,
\begin{equation}
{\bar L}_{10} = L_{10}^{(r)}(\mu)  + {144\over \pi^2}
\left[ \ln \left( {m_\pi^2\over \mu^2}\right) + 1\right]
\simeq -(6.84 \pm 0.3 ) \times 10^{-3} \ \ .
\label{10}
\end{equation}
The next two relations, $W1$ and $W2$, are respectively
the first and second Weinberg sum rules.$^{\cite{w},\cite{ed}}$  The final
sum rule, $W3$, is a formula for the $\pi^\pm$-$\pi^0$ mass splitting
in the chiral limit.$^{\cite{dgmly}}$  Although apparently containing
an arbitrary energy scale $\Lambda$, this sum rule is actually
independent of $\Lambda$ by virtue of $W2$.  For reference,
we have displayed the {\it physical} values of the nonzero entries which
appear on the right hand side of the chiral sum rules.  These
quantities have slightly shifted values in a chiral invariant world.
This point is discussed at the end of Sect.~3.

We note in passing that the current correlators
defined in Eqs.~(\ref{4a},\ref{4b}) have been the subject
of much recent attention.  Several analyses have been carried out
of hadron production in tau-lepton decay in order
to obtain a determination of $\alpha_s (m_\tau )$,
the running strong fine structure constant evaluated at the
tau mass scale.$^{\cite{qcdtau}}$
\section{\bf Data Inputs}

It is possible, in principle, to analyze
the chiral sum rules on the basis of pure
theory.  For example, in the original derivation of $W3$,
the pion electromagnetic mass difference
was estimated by using $\rho (770)$ and
$a_1 (1100)$ contributions to saturate the vector and
axialvector spectral functions.$^{\cite{dgmly}}$.
However, a more sound procedure
is to use data from tau lepton semileptonic decays
into pions and/or pion production in $e^+e^-$ annihilations.

The rate for tau decay into an even or odd number of pions
at invariant squared-energy $s$ is given by $^{\cite{tsa}}$
\begin{equation}
{d\Gamma (\tau\to \nu_\tau ~{{\rm Even}\choose {\rm Odd}})
\over ds} =
{G_\mu^2 V_{\rm ud}^2 \over 8\pi m_\tau^3}
(m_\tau^2 - s)^2\left[ (m_\tau^2 + 2s) {\rho_{\rm V}(s)\choose
\rho_{\rm A}(s)} + m_\tau^2 {0 \choose \rho_{\rm A}^{(0)}}
\right]\ ,
\label{11}
\end{equation}
and the corresponding $n\pi$ branching ratio is
\begin{equation}
B_{n\pi} = {G_\mu^2 V_{\rm ud}^2 m_\tau^2\over
8\pi\Gamma_{\tau \to {\rm all}} }I_{n\pi} \ \ {\rm with}\ \
I_{n\pi} = \int_{(nm_\pi)^2}^{m_\tau^2} ds~\left( 1 - {s\over
m_\tau^2}\right)^2
\left( 1 + {2s\over m_\tau^2}\right) \rho_{n\pi}(s) \ .
\label{12}
\end{equation}
In addition to the Particle Data Group (hereafter $PDG$)$^{\cite{rpp}}$,
the primary sources for tau decay data are the $ARGUS$
$^{\cite{arg1}}$ and $CLEO$ $^{\cite{cle}}$ detectors, via
the reaction $e^+e^-\to\tau^+\tau^-$, and the $LEP$ detectors $^{\cite{lep}}$
via the decay $Z^0\to \tau^+\tau^-$.  A review of tau physics
up to 1988 is given by Barish and Stroynowski.$^{\cite{bs}}$  Experimental
aspects of tau decay continue to be presented up to the most recent
conferences.$^{\cite{led,spa,dav}}$

At first, using $e^+e^-$ annihilation data to test the chiral sum rules
would appear to be problematic.  Although the range of energy is (at
least in principle) unlimited, the production mechanism involves the
electromagnetic current and so is generally a mixture of the needed
isospin $1$ component and an unwanted isospin $0$ component.  However,
for final states which consist of an
even number of pions it is only the isovector electromagnetic current
which contributes.  This is a consequence of the $G$-parity relation
$G_{n\pi} = (-)^n$, together with the property that any
$n\pi$ final state produced by the action of the electromagnetic
current on the vacuum must have charge conjugation $C = -1$.  Since
$G = C (-)^I$, it follows that $I=1$ for $n$ even.  Of course,
the isospin components which are measured
in $e^+ e^-$ scattering and tau decay are distinct, having
$I_3 = 0$ and $I_3 = 1 - i2$ respectively.   The corresponding
$2\pi$ and $4\pi$ states are related by
\begin{eqnarray}
T_- |\pi^+\pi^- \rangle &=& \sqrt{2} |\pi^0 \pi^- \rangle
\nonumber \\
T_- |2\pi^+ 2\pi^- \rangle &=& 2\sqrt{2} |\pi^0 \pi^+ 2\pi^- \rangle
\label{4pia} \\
T_- |\pi^+ \pi^- 2\pi^0 \rangle &=& \sqrt{2} |\pi^- 3\pi^0 \rangle +
2\sqrt{2} |\pi^+ \pi^0 2\pi^- \rangle \ \ . \nonumber
\end{eqnarray}
Extraction of the ${n\pi}$ component of
$\rho_{\rm V}(s)$ from $e^+ e^-$ data proceeds via the relation
\begin{equation}
\rho^{n\pi}_{\rm V}(s) = {1\over 16\pi^3\alpha^2}s
\sigma^{n\pi}_{{\rm I}=1}(s)\ \ .
\label{13}
\end{equation}
For the specific case of two-pion production, the
$e^+e^-$ annihilation data is often
expressed in terms of the pion electromagnetic form factor $F^{\pi\pi}(s)$
evaluated at squared-energy $s$.  In this notation, one has
\begin{equation}
\rho^{2\pi}_{\rm V}(s) = {1\over 48\pi^2}
\left( 1 - {4m_\pi^2\over s} \right)^{3/2} |F^{\pi\pi}(s)|^2 \ \ .
\label{14}
\end{equation}
A plot of the pion form factor in the timelike region appears in
Fig.~1.  Early results on pion production in $e^+e^-$ scattering
is summarized in Ref.~\cite{ww}.  Experiments at
Frascati, Orsay and Novosibirsk have continued
to supply data.$^{\cite{fra}-\cite{bar1}}$

As a whole, tau and $e^+ e^-$ data
reveal that each of the multipion contributions
rises fairly sharply from threshold to a peak value and then
falls rather more slowly.  At energies below $2$~GeV, the
role of meson resonances is significant.  Thus, the $2\pi$ contribution
has the familiar narrow resonant structure of $\rho (770)$,
the $3\pi$ modes are dominated by $a_1 (1260)$, and the $4\pi$ sector is
influenced by $\rho (1450)$ and $\rho (1700)$.  Although lacking
a detailed dynamical understanding of higher multiplicity distributions,
we can anticipate their form as a consequence of general physical
considerations.  They will occur sequentially in mass (due
to increasingly higher mass thresholds), exhibit slowly decreasing peak
values (from competition with other channels as constrained
by unitarity), and become increasingly broad (since phase space grows
with particle number).  As energy increases, one soon enters
the asymptotic domain and the sum over all modes
becomes featureless, in a manner similar to the total $e^+ e^-$
hadronic cross section.

\vspace{0.4 cm}

Let us now present a critique of the current status for various
individual contributions:
\begin{center}
{\bf Tau Lepton Properties}
\end{center}

The most recent complilation given by $PDG$
for the primary tau lepton properties of
mass ($m_\tau$), lifetime ($\tau_\tau$) and electron branching ratio
$B_e \equiv \Gamma_{\tau \to e\nu_\tau {\bar \nu}_e} /
\Gamma_{\tau \to {\rm all}}$ are
\begin{equation}
m^{PDG}_\tau \simeq 1.784\ GeV \ ,\ \
\tau^{PDG}_\tau \simeq 305~{\rm fs}\ ,\ \
B^{PDG}_e \simeq 0.1793\ .
\label{pdg}
\end{equation}
However, there have been recent downward revisions to$^{\cite{rol}}$
\begin{equation}
m_\tau \simeq 1.777\ GeV \ ,\qquad \tau_\tau \simeq 297~{\rm fs}\ ,
\qquad \ B_e \simeq 0.1771\ \ , \label{16}
\end{equation}
and these are the values that we shall use throughout our analysis.
Observe that there is a slight inconsistency between
the listed central values for $m_\tau$, $\tau_\tau$ and $B_e$ in
Eq.~(\ref{12}).  As a result, the theoretical constraint
\begin{equation}
B_e = 0.06125 {\tau_\tau\over 10^{-13}~{\rm sec}}
\label{17}
\end{equation}
is not exactly satisfied.  Thus, the relation between
$n\pi$ branching branching ratios and spectral functions in Eq.~(\ref{12})
depends on which of the following relations is assumed,
\begin{equation}
(i)\ B_{n\pi} = 4.39 {\tau_\tau\over 10^{-13}~{\rm sec}}I_{n\pi} \qquad
{\rm or}\qquad (ii)\
B_{n\pi} = 71.31~B_e~I_{n\pi}\ .
\label{19}
\end{equation}
Throughout this paper, we shall for definiteness assume that
$B_e = 0.1771$ in our numerical work.  This value is an average
of the value in Eq.~(\ref{pdg}) and the recent experimental
determination cited in Ref.~\cite{cleo}.

\begin{center}
{\bf Two-pion Component}
\end{center}

Data for $\rho_{\rm V}^{2\pi}$ comes
from the $\pi^-\pi^0$ part of one-prong $\tau$
decay$^{\cite{arg1,mar}}$ and from the
$\pi^+\pi^-$ final state in $e^+e^-$ scattering.$^{\cite{fra,zhe,bar1,bar2}}$
The consistency of tau decay and $e^+ e^-$ annihilation results
in the vicinity of the $\rho (776)$ peak has been verified by Gan
in Ref.~\cite{gan}.  We display in Fig.~2 the two-pion
spectral function $\rho_{\rm V}^{2\pi}(s)$ as inferred from the
pion form factor data of Fig.~1.  Numerical integration
of $\rho_{\rm V}^{2\pi}$ yields a $2\pi$ branching ratio of
$B_{2\pi} = 0.247$.  There are also recent determinations
of the $h^- \pi^0$ (where $h$ is a hadron) and $\pi^- \pi^0$
branching ratios by $CLEO$$^{\cite{cle}}$ and $LEP$$^{\cite{lep}}$,
\begin{eqnarray}
B_{h^-\pi^0}^{\rm CLEO} &=& 0.2483 \pm 0.0015 \pm 0.0053 \ , \nonumber \\
B_{h^-\pi^0}^{\rm LEP}  &=& 0.243 \pm 0.008 \ ,\label{20} \\
B_{\pi^-\pi^0}^{\rm CLEO} &=& 0.2435 \pm 0.0055 \ \ , \nonumber
\end{eqnarray}
where the $\pi^-\pi^0$ value is inferred by subtracting off
the $K^{*-}$ branching ratio from that of $h^- \pi^0$.
The above are in reasonable accord with the value cited by the
$PDG$$^{\cite{rpp}}$, which is based on earlier data.

All in all, the two-pion part of the vector current spectral function
is well determined.  As we shall now see, although much is known
about the three-pion and four-pion distributions, more
experimental input would be welcome.

\begin{center}
{\bf Three-pion Component}
\end{center}

We denote branching ratios for the two $3\pi$ modes in $\tau$ decay as
$B_{\pi^+2\pi^-}$ and $B_{\pi^-2\pi^0}$.
Basic isospin considerations imply the inequalities$^{\cite{gil}}$
\begin{equation}
{1\over 2} \le {B_{\pi^+2\pi^-} \over B_{3\pi}}
\le  {4\over 5} \qquad {\rm and} \qquad
{1\over 5} \le {B_{\pi^-2\pi^0} \over B_{3\pi}}
\le  {1\over 2}\ \ .
\label{21}
\end{equation}
$PDG$ lists the three-hadron branching ratios based on a number of
experiments as
\begin{equation}
B_{h^+2h^-} = 0.084 \pm 0.004 \qquad {\rm and} \qquad
B_{h^-2\pi^0} = 0.103 \pm 0.009 \ \ .
\label{21a}
\end{equation}
$CLEO$$^{\cite{cle}}$ and $LEP$$^{\cite{lep}}$ have recently
announced the results
\begin{eqnarray}
B_{h^-2\pi^0}^{\rm CLEO} &=& 0.0821 \pm 0.0015 \pm 0.0038
\pm 0.0028 \ , \nonumber \\
B_{h^-2\pi^0}^{\rm LEP} &=& 0.104 \pm 0.008 \ ,\label{22} \\
B_{h^+2h^-}^{\rm LEP} &=& 0.0949 \pm 0.0036 \pm 0.0063 \ \ . \nonumber
\end{eqnarray}
Collectively, these imply that $B_{\pi^+2\pi^-} \simeq B_{\pi^-2\pi^0}$
and indicate a total $3\pi$ branching ratio of $16\%$-$19\%$.
For definiteness, we shall assume that both modes have equal
decay rates and employ total three-pion branching ratios in this
same range.  This is in accord with the findings of Davier who,
in a review of early and recent $3\pi$ branching ratio
determinations, summarizes the current situation as$^{\cite{dav}}$
\begin{equation}
B_{\pi^+2\pi^-} = B_{\pi^-2\pi^0} = 0.0903 \pm 0.0036 \ \ ,
\label{dav}
\end{equation}
but at the same time makes the cautionary remark that `extreme care
should be exercized when using world average values for branching
ratios'.

The equality of $\Gamma_{\pi^+2\pi^-}$ and $\Gamma_{\pi^-2\pi^0}$
is expected of a final state which is dominated by the
$A_1$ resonance.  The isospin decomposition
\begin{equation}
| A_1 \rangle = {1\over \sqrt{2}} |\rho^0 \pi^- \rangle -
{1\over \sqrt{2}} |\rho^- \pi^0 \rangle
\label{rhoa}
\end{equation}
shows that the $\pi^+ 2\pi^-$ (from the first term) and the
$\pi^- 2\pi^0$ (from the second term) final states will occur
with equal probability.

As regards the spectral function $\rho_{\rm A}^{3\pi}$,
reconstruction from experiment would require $3\pi$ mass
distributions for both $\pi^+ 2\pi^-$ and $\pi^- 2\pi^0$
modes.  None of the latter exists.  However, in view of $A_1$
dominance it suffices to know just the $\pi^+ 2\pi^-$ spectrum.
We refer the reader to Ref.~\cite{bs} for histograms of the
$3\pi$ mass distribution measured some time ago by the $DELCO$,
$MAC$ and $MARK~II$ detectors.  The literature also contains an
early $ARGUS$ determination$^{\cite{arg1}}$,
corresponding to the rather small branching ratio
$B_{\pi^+2\pi^-} = 0.056 \pm 0.007$.  In this paper, we shall
employ recent $ARGUS$ data$^{\cite{arg3}}$ to determine
$\rho_{\rm A}^{3\pi}$.  From the number of counts $\Delta N$
per energy bin $\Delta E$, one can construct
the $3\pi$ spectral function via
\begin{equation}
\rho_{\rm A}^{3\pi}(s) = {m_\tau^2\over 24\pi^2 V_{\rm ud}^2}
{B_{3\pi}\over B_e}
{1\over \left( 1 - s /m_\tau^2\right)^2 \left( 1 +
2s /m_\tau^2\right)}
{1\over 2E N_{\rm tot}}{\Delta N \over \Delta E} \ \ ,
\label{rho3}
\end{equation}
where $E = \sqrt{s}$.  Upon taking $B_{3\pi} = 0.17$ rather than
the value given in Ref.~\cite{arg3} (which would imply $B_{3\pi}
= 0.13$), we obtain the spectral function shown in Fig.~3.
The large error bars near the end-point occur because one must divide
the number of counts per energy bin by a phase space factor which
vanishes at $s = m_\tau^2$.  Even so, it is clear from Fig.~3
that tau decay data is able to cover essentially all the region where
$\rho_{\rm A}^{3\pi}$ is nonvanishing.

\begin{center}
{\bf Four-pion Component}
\end{center}

In tau decay, there are two $4\pi$ modes,
$\pi^-3\pi^0$ and $\pi^+\pi^02\pi^-$.  The
corresponding four-pion final states in $e^+ e^-$ scattering
are $2\pi^+ 2\pi^-$ and $\pi^+\pi^- 2\pi^0$.  The four-pion
spectral function measured in tau decay can be decomposed as
\begin{equation}
\rho_{\rm V}^{4\pi}(s) = \rho_{\rm V}^{+--0}(s) +
\rho_{\rm V}^{-000}(s) \ \ ,
\label{4pi}
\end{equation}
where the quantities on the right-hand side are inferred
from the four-pion mass distributions in the $\pi^-3\pi^0$ and
$\pi^+\pi^02\pi^-$ modes.  It is also possible to obtain the
quantities $\rho_{\rm V}^{+--0}$ and $\rho_{\rm V}^{-000}$
from $e^+e^-\to 4\pi$ cross sections via the relations
\begin{eqnarray}
\rho_{\rm V}^{-000}(s) &=& {1\over 32 \pi^3 \alpha^2}
s\sigma_{2\pi^+ 2\pi^-} \ \ ,  \label{sp1} \\
\rho_{\rm V}^{+--0}(s) &=&  {1\over 32 \pi^3 \alpha^2 }
s(\sigma_{2\pi^+ 2\pi^-} + 2 \sigma_{\pi^+ \pi^- 2\pi^0})
\ \ .  \label{sp2}
\end{eqnarray}

The set of $\pi^-3\pi^0$ and $h^- \ge 3\pi^0$
tau decay branching ratios taken from recent conference
presentations Refs.~\cite{cle},~\cite{lep}
and the average cited by the $PDG$$^{\cite{rpp}}$
provide a reasonably consistent picture,
\begin{eqnarray}
B_{h^-3\pi^0}^{\rm CLEO}  &=& 0.0098 \pm 0.0007 \pm 0.0012
\pm 0.0003 \ , \nonumber \\
B_{h^- \ge 3\pi^0}^{\rm LEP}  &=& 0.0153 \pm 0.004 \pm 0.006 \ ,
\label{23} \\
B_{h^- \ge 3\pi^0}^{\rm PDG} &=& 0.027 \pm 0.009 \ . \nonumber
\end{eqnarray}
implying $B_{\pi^-3\pi^0} \simeq 0.01$.   To our knowledge, there
is no published {\it spectral} information from tau decay data for the
$\tau\to\pi^- 3\pi^0\nu_\tau$ mode.  However, we can obtain
$\rho_{\rm V}^{-000}(s)$ from $\sigma_{2\pi^+ 2\pi^-}$
as in Eq.~(\ref{sp1}).  The set of $2\pi^+ 2\pi^-$ cross section data
taken from Ref.~\cite{sid} (for $\sqrt{s} < 1.4$~GeV) and from
Refs.~\cite{bis},~\cite{zhe} (for $\sqrt{s} > 1.4$~GeV)
is displayed in Fig.~4.  Note that Fig.~4 clearly
demonstrates how $\rho_{\rm V}^{4\pi}$ must extend beyond $s = m_\tau^2$,
{\it i.e.} tau decay alone cannot determine the $4\pi$ spectral
function over its full range.  The resulting $\rho_{\rm V}^{-000}$
obtained in this manner is shown in Fig.~5.  From this, we
predict the $\pi^- 3\pi^0$ {\it mass spectrum} to have the form shown in
Fig.~6, where a smooth curve has been added to help
guide the eye.  Integrating this mass spectrum
leads to a branching ratio $B_{\pi^-3\pi^0} \simeq 0.009$,
in good agreement with the values of Eq.~(\ref{23}).

Turning to the larger $\pi^+2\pi^- \pi^0$ tau decay mode, we
cite the branching ratio values appearing
in Refs.~\cite{lep},~\cite{rpp},~\cite{spa},
\begin{eqnarray}
B_{h^+2h^- \ge 1\pi^0}^{\rm LEP} &=& 0.0489 \pm 0.008 \ , \nonumber\\
B_{\pi^+2\pi^- \pi^0}^{\rm ARGUS} &=& 0.054 \pm 0.004 \pm 0.005 \ , \label{24}
\\
B_{h^+2h^- \ge 1\pi^0}^{\rm PDG} &=& 0.053 \pm 0.004\ , \nonumber
\end{eqnarray}
where the $ARGUS$ value is preliminary.  Two of these values include
unknown amounts of strange particle contributions.  Allowing for such
non-$4\pi$ contributions, the $PDG$ and $LEP$ values indicate
a $\pi^+ 2\pi^- \pi^0$ branching ratio in the $0.040 - 0.050$ range.
We are aware of just one published $\pi^+ 2\pi^- \pi^0$ mass spectrum
measurement, an $ARGUS$ analysis (Ref.~\cite{arg2}) which cites the
branching ratio $B_{\pi^+2\pi^- \pi^0} = 0.042 \pm 0.005 \pm 0.009$.
This value is smaller than, although not inconsistent with,
the more recent $ARGUS$ value of Eq.~(\ref{24}).
Using an appropriate binning procedure, we can construct the full
$4\pi$ spectral function $\rho_{\rm V}^{4\pi}(s)$ over the restricted
energy region $s< m_\tau^2$ by combining the $\pi^+ 2\pi^- \pi^0$ tau
data together with the $2\pi^+ 2\pi^-$ cross sections.  The result
is shown in Fig.~7, together with an asymmetric Breit-Wigner fitting
curve (see Sect.~4).

Alternatively, one could use the combination of cross sections as in
Eq.~(\ref{sp2}) to determine the total $4\pi$ spectral function.
In principle at least, this procedure can provide the $4\pi$ spectral
function over a larger energy interval.  $\pi^+\pi^- 2\pi^0$ cross
section data taken from Ref.~\cite{kur2} (for $\sqrt{s} < 1.4$~GeV) and from
Ref.~\cite{bis} (for $\sqrt{s} > 1.4$~GeV) are displayed in Fig.~8. These
turn out to imply a $\pi^+2\pi^- \pi^0$ tau decay branching
ratio somewhat smaller than the recent determinations cited in
Eq.~\ref{24}.  For this reason, we have chosen to base our analytical
work on the determination of $\rho_{\rm V}^{4\pi}(s)$ as
shown in Fig.~7.

\begin{center}
{\bf Higher Components}
\end{center}

The five-pion component in tau decay involves the branching ratios
$B_{\pi^-4\pi^0}$, $B_{3\pi^-2\pi^+}$, and $B_{2\pi^-\pi^+2\pi^0}$.
Isospin constraints for these modes are
\begin{equation}
0 \le {B_{\pi^-4\pi^0} \over B_{5\pi}}
\le  {3\over 10}~, \qquad
0 \le  {B_{3\pi^-2\pi^+} \over B_{5\pi}}
\le {8\over 35}~, \qquad
{8\over 35} \le  {B_{2\pi^-\pi^+2\pi^0} \over B_{5\pi}}
\le 1 \ .
\label{25}
\end{equation}
At present, Refs.~\cite{cle},~\cite{rpp},~\cite{spa} provide the following
branching ratio determinations,
\begin{eqnarray}
B_{h^-4\pi^0}^{\rm CLEO}  &=& 0.0015 \pm 0.0004 \pm 0.0005
\pm 0.0001 \ , \nonumber \\
B_{3h^-2h^+}^{\rm PDG}  &=& 0.00056  \pm 0.00016 \ , \label{higher} \\
B_{2\pi^-\pi^+2\pi^0}^{\rm ARGUS}  &=& 0.0054 \pm 0.0005
\pm 0.0008 \ \ . \nonumber
\end{eqnarray}
Noting that $B_{3\pi^-2\pi^+} \le B_{3h^-2h^+}$, we find that
these values are consistent with the bounds of Eq.~(\ref{25}).
However, the isospin bounds do not imply any useful information
for disentangling $B_{3\pi^-2\pi^+}$ from $B_{3h^-2h^+}$.
To our knowledge, the above branching ratios are the only $5\pi$
data currently available. To obtain useful spectral infomration for
the $5\pi$ mode requires a substantial number of events,
{\it e.g.} as would be generated from a $\tau$ factory.

Some $6\pi$ spectral information is available from $e^+e^-$
scattering and some $6\pi$ branching ratio information is
available, but in view of the paucity of $5\pi$ data we have
not included this sector in the analysis described in this paper.

\vspace{0.4cm}

Although our statement of the chiral sum rules in Sect.~2
refers to the limit of massless quarks, the data reviewed
in this section are taken in the real world of $m_{\rm u,d} \ne 0$.
In principle, we might attempt to perform corrections on the data set
with an eye towards working in the chiral limit.  For example, it is
apparent that taking $m_\pi \to 0$ would induce minor shifts in resonance
masses and phase space.  However, we anticipate that such effects would
be of order $m_\pi^2 /\Lambda^2$ with $\Lambda \simeq 1$~GeV.  Since
such changes are much smaller than the present uncertainty in the
data, it seems most prudent not to try to model the effect of
finite $m_{\rm u,d}$ effects.  Thus, we shall use the unmodified
experimental information in our phenomenological analysis.
\section{\bf Theoretical Constraints}

Although the vector and axialvector spectral functions are not
theoretically prescribed for all values of $s$, it is
possible to place constraints on their low and high energy limits.

We begin by taking the Fourier transform of Eq.~(\ref{4a}),
\begin{eqnarray}
\delta_{ab}\Pi_{\mu\nu}^{\rm V}(q^2) &=& i\int d^4 x\ \exp^{iq\cdot x}
\langle 0|T\left( V^\mu_a (x) V^\nu_b (0)\right)|0\rangle \nonumber \\
&\equiv& \delta_{ab}\left( q_\mu q_\nu -
g_{\mu\nu}q^2 \right)\Pi_{\rm V}(q^2)\ . \label{vv}
\end{eqnarray}
and of Eq.~(\ref{4b}),
\begin{eqnarray}
\delta_{ab}\Pi_{\mu\nu}^{\rm A}(q^2) &=& i\int d^4 x\ \exp^{iq\cdot x}
\langle 0|T\left( A^\mu_a (x) A^\nu_b (0)\right)|0\rangle \nonumber \\
&\equiv& \delta_{ab}\left( q_\mu q_\nu - g_{\mu\nu}q^2 \right)\Pi_{\rm A}(q^2)
- q^\mu q^\nu \Pi_{\rm A}^{(0)}(q^2)  \ .
\label{aa}
\end{eqnarray}
In the chiral limit, the spin-$0$ axial contribution
$\Pi_{\rm A}^{(0)}(q^2)$ is given entirely by the pion pole.
The correlators $\Pi_{\rm V,A}(q^2)$ are the real parts of
analytic functions whose imaginary parts are the spectral
functions $\rho_{\rm V,A}(s)$.  Of interest to us are the
dispersion relations involving the difference of vector and
axialvector quantities,
\begin{eqnarray}
\Pi_{\rm V}(q^2) - \Pi_{\rm A}(q^2) &=& {F_\pi^2 \over q^2} +
\int_{0}^\infty ds ~{\rho_{\rm V}(s) - \rho_{\rm A}(s)\over s- q^2 -
i\epsilon} \nonumber \\
&=& {1\over q^2}\int_{0}^\infty ds \
s~{\rho_{\rm V}(s) - \rho_{\rm A}(s)\over s- q^2 - i\epsilon} \ .\label{dr}
\end{eqnarray}

In our normalization, the behavior of the individual
$\Pi_{\rm V,A}(q^2)$ to leading order at large $q^2$ is
\begin{equation}
\Pi_{\rm V,A}(q^2) \sim {1\over 8\pi^2}
\left( 1 + {\alpha_s (q^2)\over \pi}\right)
\ln\left( {\mu^2\over -q^2}\right)\ \ ,
\label{hec}
\end{equation}
where $\mu$ is the renormalization scale.  In order to
determine the {\it difference} $\Pi_{\rm V}(q^2) - \Pi_{\rm A}(q^2)$
for $q^2$ large but finite,
one must go beyond the form in Eq.~(\ref{hec}).
{}From the operator product expansion of vector and
axialvector currents, we learn that
the asymptotic dependence is ${\cal O}(q^{-6})$ and the
local operators which control this behavior are the
four-quark condensates.$^{\cite{svz},\cite{lsc}}$
In the approximation of vacuum saturation, one finds
\begin{equation}
\Pi_{\rm V}(q^2) - \Pi_{\rm A}(q^2) = {32\pi\over 9}
{\langle \sqrt{\alpha_s }{\bar q}q\rangle_0^2 \over q^6}
\left( 1 + {\alpha_s (q^2)\over 4\pi}\left[ {247\over 12} +
\ln\left( {\mu^2\over -q^2}\right)
\right]\right)\ \ .
\label{asc}
\end{equation}

It turns out that theory also predicts the low energy or threshold
behavior of the vector and axialvector correlators.  However, in this
case it is the machinary of chiral perturbation theory that is invoked
to show$^{\cite{gl}}$
\begin{equation}
\Pi_{\rm V}(q^2) - \Pi_{\rm A}(q^2) = {F_\pi^2 \over q^2} +
{1\over 48\pi^2} \left[ \ln\left( {\mu^2\over -q^2}\right)
+ {5\over 3} \right] - L_{10}^{(r)}(\mu)  \ \ ,
\label{lec}
\end{equation}
where $L_{10}^{(r)}(\mu)$ is defined in Eq.~(\ref{10}).

The above statements all involve the correlators $\Pi_{\rm V,A}$.
Similar threshold and asymptotic constraints can be placed on
the spectral functions $\rho_{\rm V,A} (s)$.  Thus,
as stated in Ref.~\cite{gl} the threshold behavior
($s\to 4m_\pi^2 $) of $\rho_{\rm V}(s)$ and $\rho_{\rm A}(s)$ is
\begin{eqnarray}
\rho_{\rm V} (s) &\sim& {1\over 48\pi^2}
\left( 1 - {4m_\pi^2\over s} \right)^{3/2} \theta(s - 4m_\pi^2) +
{\cal O}(p^2) \ \ ,\label{lesv}  \\
\rho_{\rm A} (s) &\sim& {\cal O}(p^2) \ \ . \label{lesa}
\end{eqnarray}
Later in this section, we shall sharpen the threshold result for
$\rho_{\rm A}$ by specifying the $3\pi$ threshold contribution in
more detail.  The perturbative result for the asymptotic limit
$s\to \infty$ of the individual $\rho_{\rm V,A}$ to leading order is
\begin{equation}
\rho_{\rm V,A} (s) \sim {1\over 8\pi^2}(1 + {\alpha_s (s)\over\pi}) \ \ .
\label{hes}
\end{equation}
Finally, Eq.~(\ref{asc}) determines the asymptotic form of
$\rho_{\rm V}(s) - \rho_{\rm A}(s)$ to be
\begin{equation}
\rho_{\rm V}(s) - \rho_{\rm A}(s) \sim {C\over s^3} \simeq
{8\over 9} {\alpha_s \langle \sqrt{\alpha_s }
{\bar q}q\rangle_0^2 \over s^3}\ \quad \
{\rm for\ large}\ s \ \ .
\label{ass}
\end{equation}
Note that the difference of the spectral functions is of order $\alpha_s^2$.
Our analysis of the chiral sum rules will require a numerical
value for the coefficient $C$ of the $s^{-3}$ term.  From the estimate
\begin{equation}
\langle \sqrt{\alpha_s }{\bar q}q\rangle_0^2  \simeq  (0.24~{\rm GeV})^6
\simeq 1.9\times 10^{-4}~{\rm GeV}^6
\label{est}
\end{equation}
and taking $\alpha_s \simeq 0.2$, we obtain
\begin{equation}
C \simeq 3.4\times 10^{-5}\ {\rm GeV}^6 \ \ .
\label{num}
\end{equation}
The magnitude of this quantity is obviously model
dependent and quite possibly will be modified
by future work.  However, even folding in uncertainties of the vacuum
condensates, it is clear that the coefficient of the $s^{-3}$ term is
very small, and that $\rho_V$-$\rho_A$ approaches zero very quickly at
large $s$.  Even if one imagines $\rho_V (s)$-$\rho_A (s)$ to
exhibit increasingly damped oscillations indefinitely, duality
suggests that Eq.~(\ref{ass}) captures the correct {\it average} behavior.
As we shall see, the asymptotic constraint of Eq.~(\ref{ass}) will
have significant impact in the analysis of the chiral
sum rules to come.

The results presented in this section
are well known, and collectively they represent a fairly
powerful set of conditions regarding how the chiral correlators
can behave.  Actually, additional thought can reveal
even more.  For example, let us employ the
asymptotic behavior of Eq.~(\ref{asc}) in the dispersion relation
of Eq.~(\ref{dr}).  Expansion of the dispersion integral
in powers of $q^{-2}$ yields the sum rule
\begin{equation}
{32\pi\over 9}\langle \sqrt{\alpha_s }{\bar q}q\rangle_0^2 = -
\int_{0}^\infty ds ~s^2 \left(\rho_{\rm V}(s) - \rho_{\rm A}(s)\right)\ \ .
\label{sr1}
\end{equation}
Some care must be taken to interpret this result correctly.
Observe that Eq.~(\ref{sr1}) is valid to ${\cal O}(\alpha_s )$.
Since the ${\cal O}(s^{-3})$ tail of $\rho_{\rm V}(s) -
\rho_{\rm A}(s)$ is itself of higher order in $\alpha_s$,
one must subtract it off in this relation.  Accordingly, we write
\begin{equation}
{32\pi\over 9}\langle \sqrt{\alpha_s }{\bar q}q\rangle_0^2 = -
\int_{0}^\infty ds ~s^2 \left(\rho_{\rm V}(s) - \rho_{\rm A}(s)\right)'\ \ ,
\label{sr2}
\end{equation}
where $\left(\rho_{\rm V}(s) - \rho_{\rm A}(s)\right)'$ refers to
subtraction procedure just mentioned.

Finally, let us return now to the matter of the
the threshold behavior of $\rho_{\rm A}(s)$.  It
is associated with the $3\pi$ component.  Using chiral lagrangian
methods$^{\cite{dgh}}$, we have determined that in the chiral limit the
threshold
behavior is
\begin{equation}
\langle 0|A^\mu (0)|\pi^+_{{\bf p}_+}\pi^0_{{\bf p}_0}\pi^-_{{\bf p}_-}\rangle
= {2\over 3F_\pi}(p_+^\mu + p_-^\mu - 2p_0^\mu)
- {3s_{+-} - s\over 3F_\pi s} Q^\mu
\label{3pia}
\end{equation}
where $Q\equiv p_+ + p_- + p_0$, $s \equiv Q^2$ and
$s_{+-}\equiv (p_+ + p_-)^2$.  Upon rearrangement of terms,
this can be expressed as
\begin{equation}
\langle 0|A^\mu (0)|\pi^+_{{\bf p}_+}\pi^0_{{\bf p}_0}\pi^-_{{\bf p}_-}\rangle
= {2 \over F_\pi}\left[ {Q\cdot p_0 \over Q^2}Q^\mu - p_0^\mu \right]\ \ .
\label{3pib}
\end{equation}
Observe that this obeys $\langle 0| \partial_\mu A^\mu (0)
|3\pi\rangle = 0$, as must be the case since it is only
the spin-one part of the axial current which contributes here.
The threshold behavior of $\rho^{3\pi}_{\rm A}(s)$
can then be read off from the general
form of Eq.~(\ref{non}) by first squaring Eq.~(\ref{3pib})
and integrating over $3\pi$ phase space,
\begin{equation}
\rho^{3\pi}_{\rm A}(s) = {s\over 96\pi (4\pi F_\pi )^2} + \dots \ \ .
\label{3pic}
\end{equation}
In practice, however, the very low s behavior for the $3\pi$ state
turns out to be less important in the chiral sum rules than the
higher energy effect of the $A1$ resonance.
\section{\bf Empirical Determinations of Chiral Sum Rules}

Before discussing our own methodology, we wish to take note of
two interesting works involving aspects of chiral sum rules.
In the earlier of these$^{\cite{ps}}$,
spectral functions based on $2\pi$, $3\pi$ and
$4\pi$ data extracted from tau decay are used to
study the sum rules $W1$, $W2$ and $W3$.  The associated
spectral integrals are studied as a function of cutoff $s_0$
for $s_0 \le 2.5$~GeV$^2$.  However, the $3\pi$ analysis was taken
from Ref.~\cite{arg1} data, which as we have seen is not in agreement
with a large number of more recent branching ratio measurements.
In addition, the $2\pi$ spectral function was too small
in the vicinity of the $\rho (770)$ peak by roughly a
factor of $2$.  Interestingly, these two features combined to make the
sum rules appear reasonably in agreement with expectations, although this
result was fortuitous.

In the analysis of Ref.~\cite{pt}, $2\pi$ and $4\pi$ data are
inferred from $e^+e^-$ scattering, but the $3\pi$ data
again comes from Ref.~\cite{arg1}.  In fitting these components, use is
made of Breit-Wigner resonance forms with energy-dependent decay
widths.  The particular form of energy dependence is taken from
Lorentz-invariant phase space.  The $5\pi$ and higher components
are parameterized to give rise to the asymptotic behavior
\begin{equation}
\rho_{\rm {V,A}}(s) \sim {5\over 32\pi^2}
\qquad {\rm and} \qquad
\rho_{\rm V} - \rho_{\rm A} \sim {\cal O}(s^{-1})\ \ .
\label{40}
\end{equation}
This disagrees with the chiral and operator product expansion results of
Eq.~(\ref{hes}) and Eq.~(\ref{ass}) respectively.  The specific form
in Ref.~\cite{arg1} used for the higher components is chosen to fit
the sum rule $W1$ exactly and the sum rule $W0$ is evaluated in terms of
the fit.

The summary given in Sect.~III of available data demonstrates
that there is hope for successfully
extracting much about the spectral functions $\rho_{\rm V,A}(s)$ from
experiment, but that our knowledge of them will always be
limited.  In view of the current database, we have decided
that for the purpose of testing the chiral sum rules,
it is most prudent to take the empirical $2\pi$,
$3\pi$ and $4\pi$ modes explicitly into account, and to
treat all higher components according to some reasonable
prescription.  We have followed two distinct approaches in doing so:
\begin{enumerate}
\item {\it Numerical:}~ We ensure that the empirical databases for
$\rho_{\rm V,A}(s)$ evolve smoothly in the variable $s$ to the correct
asymptotic limits by generating smooth curves separately for $\rho_{\rm V}$
and $\rho_{\rm A}$ which pass through the experimental data sets at
low energy and which satisfy the asymptotic limits, while reproducing the
four chiral sum rules.

\item {\it Analytical:}~ This actually encompasses a class of
fits to the difference of spectral functions
$\rho_{\rm V} - \rho_{\rm A}$ in which all contributions higher than
the four-pion sector are lumped into a single theoretical term.
A convenient method for constructing the spectral functions this
way is to start with a delta function form, then introduce
finite-widths via Breit-Wigner representations, and finally modify
these to a more realistic asymmetric form.
\end{enumerate}
Let us consider each possibility in turn.

\begin{center}
{\bf Numerical Representation}
\end{center}

Surely the simplest method for generating acceptable global
versions of $\rho_{\rm V,A}$ is to numerically smoothly join
the low energy empirical data with asymptotic theoretical information.
A reasonable region for matching the two occurs at about
$s\simeq 4-5$~GeV$^2$.  As we have already seen, the constraint
of Eq.~(\ref{hes}) reveals that the spectral functions approach a nonzero
constant at infinite energy, with an additive correction factor
proportional to $\alpha_s (s)$.  In general, we expect the large $s$ behavior
\begin{equation}
\rho_{\rm V,A}(s) = \rho_0^{\rm V,A} + {\rho_1^{\rm V,A} \over s}
+ {\rho_2^{\rm V,A} \over s^2} + \dots \ \ ,
\end{equation}
We can obtain a determination of the dominant power correction in the
large $s$ limit as follows.  Using an operator product expansion,
Braaten, Narison and Pich have displayed
the structure of the correlators $\Pi_{\rm V,A}$ for Euclidean
momenta $-Q^2$ in Appendix~A of their paper$^{\cite{qcdtau}}$,
\begin{equation}
\Pi_{\rm V,A}(-Q^2) = a_0^{\rm V,A}(Q) + {a_1^{\rm V,A}(Q) \over Q^2}
+ {a_2^{\rm V,A}(Q) \over Q^4} + \dots \ \ ,
\label{bnp}
\end{equation}
where $s = -Q^2$ and each of the $a_n^{\rm V,A}(Q)$ is expandable in
powers of $\alpha_s (Q)$.  We require the imaginary part of this
expression, analytically continued to timelike momenta.  Since the
${\cal O}(Q^{-2})$ contribution to $\Pi_{\rm V,A}$ turns out to be
proportional to quark mass, the coefficient $\rho_1^{\rm V,A}(s)$ vanishes
in the chiral limit.  Thus, the first nonvanishing subleading contribution
is the ${\cal O}(Q^{-4})$ component, from which we extract the result
\begin{equation}
\rho_2^{\rm V,A}(s) \sim {11\over 192\pi^2} \alpha_s^2 \langle
{\alpha_s \over \pi} G^2 \rangle_0 \qquad {\rm for\ large}\ s
\ \ .
\end{equation}

Let us estimate the magnitude of this quantity at $s = 5$~GeV$^2$.
The strong fine structure constant is determined in terms of an
assumed value for the $QCD$ scale parameter,
\begin{equation}
\Lambda_{\rm QCD} = 150~{\rm MeV} \ \ \Rightarrow \ \
\alpha_s (5~{\rm GeV}^2) \simeq 0.26
\label{als}
\end{equation}
and the gluon condensate from phenomenological applications of $QCD$ sum rules,
\begin{equation}
\langle {\alpha_s \over \pi} G^2 \rangle_0
= (0.02 \pm 0.01)~{\rm GeV}^4 \ \ .
\label{glue}
\end{equation}
Altogether, these values imply that the ${\cal O}(s^{-2})$ component to
$\rho_{\rm V,A}$ has a tiny coefficient,
\begin{equation}
\rho_2^{\rm V,A} \simeq 0.9\times 10^{-5}~{\rm GeV}^4\ \ .
\end{equation}
As a result, even at the modest energy $s \simeq 5$~GeV$^2$ the
${\cal O}(s^{-2})$ term has a negligible effect.  Of course, this common
addition to $\rho_{\rm V}$ and $\rho_{\rm A}$ will cancel when the difference
is taken and hence will not contribute to the chiral sum rules. In addition,
the
difference of the spectral functions was chosen to be compatible with the
asymptotic constraint given in the previous section. Again the magnitude
of this contribution is so small that it is essentially irrelevant at
modest energies. Thus while we have made an effort to generate spectral
functions with the right high energy behavior, the precise value of the
high energy terms is not important since their numerical size is quite small.

Let us summarize our procedure at this point.  We have generated numerical
representations for $\rho_{\rm V}$ and $\rho_{\rm A}$ which fit all available
data on multipion production, and which are compatible with the theoretical
behavior expected at high energy, and which when integrated yield the
correct experimental values for the four chiral sum rules. Although highly
constrained at low and high energies, the spectral functions have a modest
uncertainty in the $s = 2-4 GeV^2$ range, and this was exploited in order to
precisely duplicate the expected values of the sum rules $W0, W1, W2, W3$.
Of course, both this numerical procedure and the analytic one to follow
are subject to the choice of input values.  Given the uncertainties
in especially the $3\pi$ and $4\pi$ branching ratios, we have explored a
variety of possibilities.  Corresponding to the input set
\begin{equation}
B_{2\pi} = 0.240~, \qquad B_{3\pi} = 0.165~, \qquad
B_{4\pi} = 0.048 \ ,
\end{equation}
curves for $\rho_{\rm V}$, $\rho_{\rm A}$ and
$\rho_{\rm V}$-$\rho_{\rm A}$ are displayed respectively in Figs.~9-11.

\begin{center}
{\bf Analytical Representations}
\end{center}
\vspace {0.2cm}

(a) {\bf Delta Function}

\vspace {0.15cm}

In the delta function description, the spectral functions are
\begin{equation}
\rho_{\rm V}(s) - \rho_{\rm A}(s) =
\sum_{k=1}^4~(-)^{k+1} F_k^2 \delta(s - m_k^2) \ \ .
\label{del1}
\end{equation}
This representation of the spectral functions,
although crude, provides a nice pedagogical
example with which to organize one's thoughts.  We begin by
noting that there are $2$ parameters per contribution,
a mass $m$ and a coupling $F$.  In this approach, the four
sum rules reduce to
\begin{eqnarray}
W0 &=& \sum_{k=1}^4~(-)^{k+1} {F_k^2 \over m_k^2}\ , \label{33}\\
W1 &=& \sum_{k=1}^4~(-)^{k+1} F_k^2 = F_\pi^2 \ , \label{34}\\
W2 &=& \sum_{k=1}^4~(-)^{k+1} m_k^2~F_k^2 = 0\ , \label{35}\\
W3 &=& \sum_{k=1}^4~(-)^{k+1} m_k^2~F_k^2~\ln {m_k^2} \ , \label{36}
\end{eqnarray}
and the tau decay branching ratios become
\begin{equation}
B_{k} = 71.62~B_e~F_k^2~\left(1 - {m_k^2\over m_\tau^2}\right)^2
\left( 1 + {2m_k^2\over m_\tau^2}\right)\qquad (k=1,2,3)\ .
\label{37}
\end{equation}
We can specify the mass parameters
$m_{1,2,3}$ from the observed $n\pi$ ($n=2,3,4$) distributions
and the coupling strengths $F_{1,2,3}$ from the observed
branching ratios ({\it cf.} Eq.~(\ref{12})).  Even in the
extreme narrow width approximation of the delta function representation,
this step turns out to be a surprisingly accurate one, {\it e.g.}
finite width effects modify the branching ratio relations
by only a few percent.

This leaves the problem of determining
the coupling $F_4$ and mass $m_4$.  It seems
reasonable to require that the sum rules
$W1$ and $W2$ be obeyed exactly ({\it cf.} Eq.~(\ref{34})
and Eq.~(\ref{35})).  Thus we determine $F_4$ from Eq.~(\ref{34})
and $m_4$ from Eq.~(\ref{35}).  Doing so leaves the
remaining sum rules $W0$ and $W3$ as predictions.
\vspace{0.2cm}

(b) {\bf Breit-Wigner}

\vspace{0.15cm}
In the finite width Breit-Wigner ($BW$ hereafter) extension of
Eq.~(\ref{del1}), the spectral functions are represented by
\begin{equation}
\rho_{\rm V}(s) - \rho_{\rm A}(s) =
\sum_{k=1}^4~{(-)^{k+1}  \over \pi}{F_k^2 m_k \Gamma_k  \over
(s - m_k^2)^2 + (m_k\Gamma_k)^2 } \ \ .
\label{bw1}
\end{equation}
Each contribution still has mass and coupling parameters $m_k$
and $F_k$, but now in addition a width $\Gamma_k$.  In the most general
Breit-Wigner resonance form, decay widths are taken to be
energy-dependent, {\it e.g.} the $2\pi$ widths have the
${\cal O}({\bf p}^3)$ threshold behavior to reflect the $P$-wave
nature of the $2\pi$ system.  However, in order to maintain
a simplicity of description and the ability to represent each
$n\pi$ component in analytic form, we shall employ
energy-independent widths throughout.

An advantage of the delta-function approach was the ability
to express all the sum rules and the tau branching ratios
in elementary form.  This is still true in
the $BW$ approximation for the sum rules $W0$, $W1$, $W2$ and the
tau branching ratios.  For example, working with a
generic Breit-Wigner spectral function,
\begin{equation}
\rho_{\rm BW} (s) = {F^2 m^2 r \over (s - m^2)^2 + (m^2 \pi r)^2 } \ \ .
\label{bw2}
\end{equation}
where $r \equiv \Gamma /(\pi m) $ is an expansion parameter for
finite-width effects, we can evaluate integrals such as
\begin{eqnarray}
I_{\rm BW}^{(0)} &\equiv& \int_{s_0}^\Lambda ds~\rho_{\rm BW}(s) \nonumber \\
&=& F^2 \left[ 1 - {1\over \pi}\arctan {\pi r\over (\Lambda /m^2) - 1}
- {1\over \pi}\arctan {\pi r\over 1 - (s_0 /m^2)}\right]\ \ ,
\label{bw3}
\end{eqnarray}
and
\begin{eqnarray}
I_{\rm BW}^{(1)} &\equiv& \int_{s_0}^\Lambda ds~s\rho_{\rm BW}(s) \nonumber \\
&=& m^2 I_{\rm BW}^{(0)} + F^2 r\left[ \ln{\Lambda\over m} +
0.5  \ln \left(
{\left(1 - m^2 /\Lambda\right)^2 + m^4\pi^2 r^2 /\Lambda^2 \over
\left(1 - s_0/m^2\right)^2 + \pi^2 r^2 }\right) \right] \ \ .
\label{bw4}
\end{eqnarray}
By passing to the limit $r = 0$ of zero decay width, we regain
the results of the delta function model.

Analogous to the procedure used in the delta function model,
parameters for the first $3$ $BW$ poles are obtained by fitting to
experimental data.  In passing, we note that since a given
spectral contribution has the asymptotic behavior ${\cal O}(s^{-2})$,
the {\it individual} $BW$ spectral integrals for $W2$ and $W3$ are
divergent.  The numerator of the fourth pole is fixed by demanding
that the ${\cal O}(s^{-2})$ asymptotic term in
$\rho_{\rm V}$-$\rho_{\rm A}$ vanish.  This implies
\begin{equation}
\sum_{k=1}^4~(-)^{k+1} F_k^2 m_k \Gamma_k = 0 \ .
\label{39}
\end{equation}
By virtue of this relation, the asymptotic behavior of
the vector and axialvector spectral functions becomes
\begin{equation}
\rho_{\rm V}(s) - \rho_{\rm A}(s) = {C_{\rm BW}\over s^3} + {\cal O}(s^4)
\label{s3}
\end{equation}
where
\begin{equation}
C_{\rm BW} = {2\over \pi} \sum_{k=1}^4~(-)^{k+1} F_k^2 m_k^3 \Gamma_k \ \ .
\label{coef}
\end{equation}
The constant $C_{\rm BW}$ may be considered either as a quantity to
be fixed by Eq.~(\ref{ass}) or as a prediction of the analysis.

We find it hard to see how any analytical study of the sum rules could
succeed without incorporating the above features or something similar.
In our approach, all the chiral sum rules are convergent even though
individual pole contributions may diverge.  Moreover, the
condition given by Eq.~(\ref{39}) ensures that our description
has the smoothness in energy expected from duality.
\vspace{0.2cm}

(c) {\bf Asymmetric Breit-Wigner}
\vspace{0.15cm}

Although having the virtue of simplicity, the
representation of Eq.~(\ref{bw1}) is deficient in several
important respects.  A Breit-Wigner form is
symmetric about its resonant energy whereas the $n\pi$
contributions to $\rho_{\rm V,A}$ exhibit asymmetric
bumps.  Besides, the $BW$ contributions extend to
energies lying below thresholds which characterize the
various $n\pi$ components.

An improved treatment can be realized in a variety of ways.
One simple parameterization which treats the various components
uniformly, yields a reasonable fit to data and allows us to
maintain analytic control is
\begin{equation}
\rho_{\rm V}(s) - \rho_{\rm A}(s) =
\sum_{k=1}^4~(-)^{k+1} {P_k(s)F_k^2 m_k \Gamma_k \over
(s - m_k^2)^2 + (m_k\Gamma_k)^2 } \ \ .
\label{39a}
\end{equation}
The functions $P_k(s)$ are polynomials
\begin{equation}
P_k (s) = \left( 1 - {s^2_k\over s^2}\right)^{n_k}\ \ ,
\label{39b}
\end{equation}
each containing two parameters, a threshold energy $s_k$ and an
integer-valued exponent $n_k = 1,2,\dots$.  Although the choice
$n_k = 1$ is the simplest one, it yields an $n\pi$ spectral
function which rises linearly just above threshold.  This does not
appear to provide an adequate fit to the data, and thus we
have used $n_k = 2 \ (k = 1 \ldots 4)$.  A fit of this type to the
$3\pi$ spectral function appears in Fig.~12.

Introduction of the polynomials $P_k (s)$ will clearly lead
to integrals involving inverse moments of Breit-Wigner forms,
\begin{equation}
I_{\rm BW}^{(-n)} =
\int_{s_0}^\infty ds\ {\rho_{\rm BW}(s) \over s^n} \ \ .
\label{rr1}
\end{equation}
Although it is straightforward to evaluate such integrals analytically,
the resulting expressions can be quite cumbersome.  In practice,
it is more efficient to employ the recursion relation
\begin{equation}
I_{\rm BW}^{(-n)} = {1\over m^4 (1 + \pi^2 r^2)} \left(
2m^2 I_{\rm BW}^{(-n+1)} - \ I_{\rm BW}^{(-n+2)}
+ {s_0^{-n + 1} \over n - 1} \right) \ \ .
\label{rr2}
\end{equation}
Upon applying relations of this type, we have carried out the calculational
program described earlier in this section.  As stated earlier, because of
the uncertainty in the experimental $3\pi$ and $4\pi$ contributions,
we have performed the analysis for several different sets of tau lepton
branching ratios.  Typical results are shown in Table~1, where we
display both input values (parameters for the $2\pi$, $3\pi$ and $4\pi$
asymmetric $BW$ poles and the associated branching ratios) and results
(parameters for the $4^{th}$ pole, the coefficient $C$ and the values for
$W0$ and $W3$).$^{\cite{demur}}$  We have purposely exhibited two solutions
($\#1$ and $\#2$) to show that it is possible to fit the sum rules yet
not obtain an acceptable value of $C$.  The graph of
$\rho_{\rm V}$-$\rho_{\rm A}$ corresponding to solution $\# 3$ of
Table~1 is displayed in Fig.~13.  The overall appearance of
this curve clearly mimics the one in Fig.~11
which is based on the numerical approach. We view the fact that
the different methods generate rather similar spectral functions as an
indication that there is not great freedom in $\rho_{\rm V} -\rho_{\rm A}$
once the theoretical and experimental constraints are imposed.
\section{\bf Concluding Remarks}

What we have done in this paper is to suggest a procedure
for comparing four well-known sum rules of
chiral symmetry with data from the real world of
experiment.  In addition, we have phenomenologically
constructed the vector and axialvector spectral functions
to the extent possible by using the full collection of available
tau decay and $e^+ e^-$ cross section data.  To be specific, the data
which constrains the spectral functions includes the $e^+ e^- \to \pi^+
\pi^- , 2\pi^+2\pi^-$ and $\pi^+\pi^-2\pi^0$ cross sections, the
$\tau \to 2\pi, 3\pi$ and $4\pi$ branching ratios, and the energy
spectra for $\pi^-\pi^0 , 2\pi^-\pi^+$ and $2\pi^-\pi^+\pi^0$ final states in
$\tau$ decay. Theoretical constraints include chiral symmetry at low energy,
isospin relations for handling the data and the operator product
expansion of $QCD$ at high energy. It is clear to us that this activity will
be repeated by ourselves or by others in the future.
That is, we (perhaps optimistically) anticipate the
emergence of improved data which will provide a
yet more reliable foundation upon which to base the
phenomenology.  However, we feel that our results are the
the best determination of the spectral functions that can be made at
this time.

As regards the data, we urge that efforts be made to
improve the determinations of the $3\pi$ and $4\pi$ contributions
to the spectral functions.  The $3\pi$ component can only be
inferred from the tau decay hadronic distribution.  In this
paper, we were fortunate to have access to the recent $ARGUS$
determination of $\rho_{\rm A}^{3\pi}$.  Cross checks are always
welcome, so we urge that data from tau production at both $LEP$
and $CESR$ be analyzed to extract the spectral function
$\rho_{\rm A}^{3\pi}$.  In addition, the need for an
improved $4\pi$ determination in $e^+ e^-$ cross section data
is especially acute for the $\pi^+\pi^- 2\pi^0$ final state.
Additional information on the $4\pi$ tau decay modes
($\pi^+\pi^0 2\pi^-$ or $\pi^- 3\pi^0$) would also be welcome.

In the latter part of our paper, we addressed the question of whether
experimental data is consistent with the chiral sum rules.  On the basis
of our study, we conclude that existing data is indeed consistent
with the chiral sum rules.  It is important to not
misinterpret this remark.  Of course, since physical data will always
be less than perfect, it is not possible to claim `proof' of
validity for the set of sum rules.  It is evident to us that to
insist on such proof would be foolhardy.  However, given the number of
constraints on the spectral functions, it is far from trivial that all
of the chiral sum rules can be satisfied. Agreement at the level
we obtained is an affirmation of the subtle and complex
theoretical intuition (involving chiral symmetry, dispersion
relations, the operator product expansion and the asymptotic
behavior of $QCD$) that leads to the sum rules.

\vspace{1.0 cm}

The research described in this paper was supported in part by
the National Science Foundation.  We wish to thank K. K. Gan,
J. J. Gomez, F. Le Diberder, and B. Spaan for helpful correspondence and/or
discussions regarding experimental aspects of spectral function
determinations.  We also thank A. P\'erez and A. Pich for useful
observations.

\eject
\phantom{xxxx}\vspace{0.5in}
\begin{center}
\begin{tabular}{||c|l|l|l||} \hline\hline
{\bf Trial} & {\bf \#1} & {\bf \#2} & {\bf \#3} \\ \hline\hline
\multicolumn{4}{||c||}{{\bf Two-pion Inputs}} \\ \hline
$M_1$ & $0.763$ &  $0.763$ &  $0.763$  \\
$f_1$ & $0.157$ & $0.157$ & $0.157$ \\
$\Gamma_1$ & $0.123$ & $0.123$ & $0.123$ \\
$s_1$ & $4 m_\pi^2$ & $4 m_\pi^2$ & $4 m_\pi^2$ \\
$B_{2\pi}$ & $0.243$ & $0.243$ & $0.243$ \\ \hline\hline
\multicolumn{4}{||c||}{\bf Three-pion Inputs}  \\ \hline
$M_2$ & $1.117$ &  $1.117$ &  $1.117$  \\
$f_2$ & $0.244$ & $0.234$ & $0.250$ \\
$\Gamma_2$ & $0.470$ & $0.470$ & $0.470$ \\
$s_2$ & $0.500$ & $0.510$ & $0.550$ \\
$B_{3\pi}$ & $0.185$ & $0.168$ & $0.176$ \\ \hline\hline
\multicolumn{4}{||c||}{\bf Four-pion Inputs}  \\ \hline
$M_3$ & $1.500$ &  $1.500$ &  $1.490$  \\
$f_3$ & $0.192$ & $0.188$ & $0.215$ \\
$\Gamma_3$ & $0.564$ & $0.700$ & $0.765$ \\
$s_3$ & $0.700$ & $0.710$ & $0.690$ \\
$B_{4\pi}$ & $0.048$ & $0.041$ & $0.053$ \\ \hline\hline
\multicolumn{4}{||c||}{\bf Output Values}  \\ \hline
$M_4$ & $2.288$ &  $2.055$ &  $1.869$\\
$f_4$ & $0.068$ & $0.083$ & $0.116$ \\
$\Gamma_4$ & $0.221$ & $0.750$ & $0.873$ \\
$B_{n>4\pi}$ & $0.0001$ & $0.0015$ & $0.0046$ \\
$W0$ & $0.0261$ & $0.0266$ & $0.0268$ \\
$W3$ & -$0.0062$ & -$0.0062$ & -$0.0062$ \\
$C$ & $0.013$ & $0.003$ & $3.8\times 10^{-5}$ \\ \hline\hline
\end{tabular}
\end{center}
\vspace{0.5 in}
\begin{center}
Table I: {Asymmetric Breit-Wigner Representation.}  \\
{The energy unit is GeV.} \\
\end{center}
\vfill \eject
\begin{center}{\bf\large Figure Captions}
\end{center}
\vspace{0.3cm}
\begin{flushleft}
Fig.~1 \hspace{0.2cm} Timelike pion form factor \\
\vspace{0.3cm}
Fig.~2 \hspace{0.2cm} $2\pi$ vector spectral function \\
\vspace{0.3cm}
Fig.~3 \hspace{0.2cm} $3\pi$ axialvector spectral function \\
\vspace{0.3cm}
Fig.~4 \hspace{0.2cm} Cross section for $e^+ e^- \to 2\pi^+ 2\pi^-$ \\
\vspace{0.3cm}
Fig.~5 \hspace{0.2cm} $\rho_{\rm V}^{-000}$ inferred from
$e^+e^-$ scattering \\
\vspace{0.3cm}
Fig.~6 \hspace{0.2cm} Predicted $4\pi$ mass spectrum in
$\tau\to\pi^- 3\pi^0 \nu_\tau$  \\
\vspace{0.3cm}
Fig.~7 \hspace{0.2cm} $\rho_{\rm V}^{4\pi}$ from tau decay and
$e^+ e^-$ scattering \\
\vspace{0.3cm}
Fig.~8 \hspace{0.2cm} Cross section for $e^+ e^- \to \pi^+ \pi^- 2\pi^0$ \\
\vspace{0.3cm}
Fig.~9 \hspace{0.2cm} Numerical fit to $\rho_{\rm V}$ \\
\vspace{0.3cm}
Fig.~10 \hspace{0.2cm} Numerical fit to $\rho_{\rm A}$ \\
\vspace{0.3cm}
Fig.~11 \hspace{0.2cm} Numerical fit to $\rho_{\rm V} - \rho_{\rm A}$ \\
\vspace{0.3cm}
Fig.~12 \hspace{0.2cm} Fit of asymmetric Breit-Wigner solution to $3\pi$
spectral function \\
\vspace{0.3cm}
Fig.~13 \hspace{0.2cm} Fit of asymmetric Breit-Wigner solution $\# 3$ to
$\rho_{\rm V} - \rho_{\rm A}$ \\
\end{flushleft}
\vfill \eject

\begin{thebibliography}{99}
\bibitem{anat} There turns out to be another sum rule of
physical interest, related to the chiral limit of $K$-to-$\pi$
matrix elements of a $\Delta S = 1$ nonleptonic operator.
This will be described in a separate publication.
\bibitem{demur} For completeness, the tau lepton branching ratio for
all $n\pi$ modes with $n>4$ is also included.  Although the simplicity
of our description precludes taking such values too literally, the
results are seen to be reassuringly small.
\end{thebibliography}
\end{document}